\documentclass[aps,12pt,superscriptaddress,preprint,tightenlines,floats,%
               nofootinbib,preprintnumbers]{revtex4}
\usepackage{amsmath} % Allows scaleable \text
\usepackage{epsfig}  % For eps figures
\usepackage{dcolumn} % Allows d in tables to specify aligning decimal point
\makeatletter
\@ifundefined{onlinecite}{}{} % for revtex4
\@ifundefined{inlinecite}{}{}
{\catcode`\%=12 \gdef\percent{%}}
\makeatother

\newcommand\ds{\displaystyle}
\newcommand\etal{{\it et al.\spacefactor1000}}
\newcommand\ibid{{\it ibid.\spacefactor1000}}

\newcommand\MSbar{\ensuremath{\text{$\overline{\rm MS}$}}}
\newcommand\lhs{\hbox{l.h.s.}}
\newcommand\rhs{\hbox{r.h.s.}}
\newcommand\ie{\hbox{\it i.e.\/}}
\newcommand\eg{\hbox{\it e.g.\/}}
\def\half{{\textstyle{\frac{1}{2}}}}
\def\quarter{{\textstyle{\frac{1}{4}}}}
\def\sixth{{\textstyle{\frac{1}{6}}}}
\def\CC{{\ensuremath{\cal C}}}
\def\CD{{\ensuremath{\cal D}}}
\def\CF{{\ensuremath{\cal F}}}
\def\CL{{\ensuremath{\cal L}}}
\def\CO{{\ensuremath{\cal O}}}
\def\CS{{\ensuremath{\cal S}}}
\def\CT{{\ensuremath{\cal T}}}
\def\CW{{\ensuremath{\cal W}}}
\def\gam#1{\ensuremath{\overline{(\gamma_{#1}\otimes I)} }}
\def\ixi#1{\ensuremath{\overline{(I\otimes\xi_{#1})} }}
\def\sfno#1#2{\ensuremath{\overline{(\gamma_{#1}\otimes\xi_{#2})}}}
\def\semitimes{\ensuremath{\mathrel>\joinrel\mathrel\triangleleft}}
\def\slash#1{\ensuremath{\mbox{$\not \!\! #1$}}}
\def\MeV{{\ensuremath{\mathop{\rm MeV}\nolimits}}}
\def\GeV{{\ensuremath{\mathop{\rm GeV}\nolimits}}}
\def\Tr{{\ensuremath{\mathop{\sf Tr}}}}
\def\Re{{\ensuremath{\mathop{\sf Re}}}}
\def\bar{\overline}
\def\hat{\widehat}
\def\tilde{\widetilde}
\def\gsim{{\mathrel{\raise2pt\hbox to 8pt{\raise -5pt\hbox{$\sim$}\hss{$>$}}}}}
\def\rsim{{\mathrel{\raise2pt\hbox to 8pt{\raise -5pt\hbox{$\sim$}\hss{$>$}}}}}
\def\lsim{{\mathrel{\raise2pt\hbox to 8pt{\raise -5pt\hbox{$\sim$}\hss{$<$}}}}}

\newcommand\overleft[1]{\ensuremath{\mathord{\mathop{#1}\limits^\leftarrow}}}
\newcommand\overright[1]{\ensuremath{\mathord{\mathop{#1}\limits^\rightarrow}}}
\newcommand\overleftright[1]{\ensuremath{\mathord{\mathop{#1}\limits^\leftrightarrow}}}
\newcommand\psibar{{\ensuremath{\mathord{\overline\psi}}}}
\newcommand\slashnext[1]{\mathpalette{\bgroup\let\style=}
                                     {\setbox0=\hbox{$\style #1$}%
                                      \setbox2=\hbox to\wd0{\hss$\style/$\hss}%
                                      \wd2=0pt\dp2=0pt\box2\box0\egroup}}
\newcommand\onelink{%
   {\mathchoice{\mathord{\rm\hbox{\the\textfont\fam 1-link}}}%
               {\mathord{\rm\hbox{\the\textfont\fam 1-link}}}%
               {\mathord{\rm\hbox{\the\scriptfont\fam 1-link}}}%
               {\mathord{\rm\hbox{\the\scriptscriptfont\fam 1-link}}}}}
\newcommand\Dslash{{\ensuremath{\mathord{\slashnext D}}}}
\newcommand\Wilson{{\ensuremath{\mathord{\cal W}}}}

\newcommand\lc[1]{\lowercase{#1}}

\usepackage{hyperref}

\begin{document}

% declarations for front matter
\preprint{LA-UR-02-7498}
%% \preprint{UW/PT-00-16}
\pacs{????}
\title{Status of $B_K$ from Lattice QCD\vspace*{1 cm}}

\author{Rajan Gupta}
\email{rajan@lanl.gov}\homepage{http://t8web.lanl.gov/t8/people/rajan/}
\affiliation{Theoretical Division, Los Alamos National Lab, Los Alamos,
         New Mexico 87545, USA\vspace*{5 cm}
	    }

\begin{abstract}
A brief review of lattice calculations of the bag parameter $B_K$
relevant for understanding indirect CP violation in the neutral kaon
sector is given. A status report on current state-of-the-art
calculations is presented as well as a discussion of the value of
$B_K$ exported to phenomenologists. This is a condensed and updated
version of the review presented at the CKM Unitarity Triangle
Workshop held at CERN during Feburary 13-16, 2002
(http://ckm-workshop.web.cern.ch/ckm-workshop/). 

\end{abstract}

% typeset front matter (including abstract)
\maketitle

\section{Introduction}
\label{sec:intro}

\begin{figure}[tbp]  % 1
\begin{center}
\epsfxsize=0.7\hsize 
\epsfbox{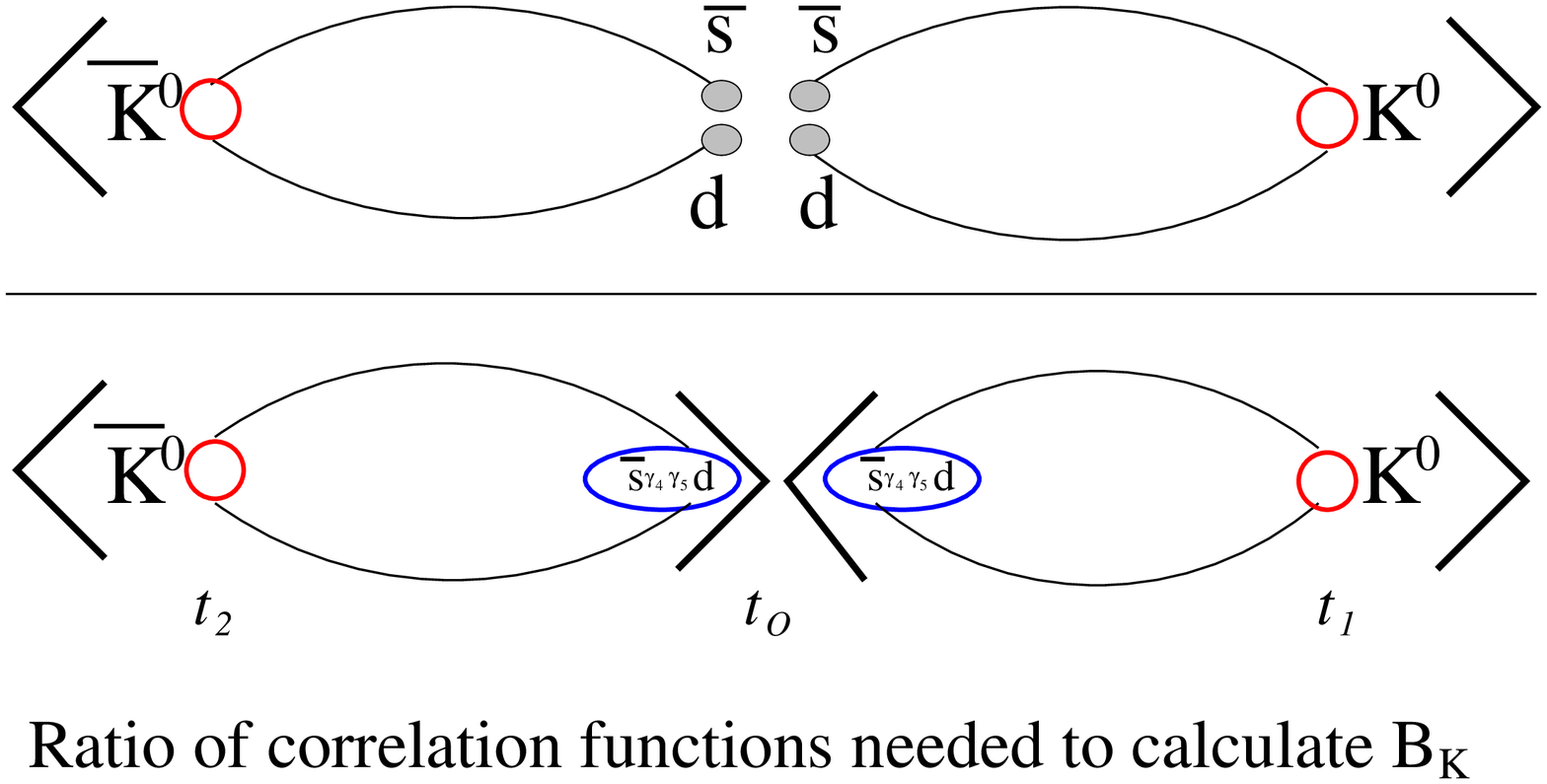}
\end{center}
\caption{}
\label{fig:bkfig}
\end{figure}

The most commonly used method to calculate the matrix element 
$\left\langle \overline{K^0} \mathbin\vert 
     Z\ (\bar s d)_{V-A} (\bar s d)_{V-A}(\mu)
        \mathbin\vert {K^0}   \right\rangle$ 
is to evaluate the three point correlation function shown in Fig.~\ref{fig:bkfig}.
This corresponds to creating a $ {K^0}$ at some time $t_1$ using a
zero-momentum source; allowing it to propagate for time $t_{\cal
O}-t_1$ to isolate the lowest state; inserting the four-fermion
operator at time $t_{\cal O}$ to convert the ${K^0}$ to a
$\overline{K^0}$; and finally allowing the $\overline{K^0}$ to
propagate for long time $t_2 - t_{\cal O}$. To cancel the $K^0$
($\overline{K^0}$) source normalization at times $t_1$ and $t_2$ and
the time evolution factors $e^{-E_K t}$ for times $t_2 - t_{\cal O}$
and $t_{\cal O}-t_1$ it is customary to divide this three-point
function by the product of two 2-point functions as shown in Fig
1. If, in the 2-point functions, the bilinear operator used to
annihilate (create) the ${K^0}$ ($\overline{K^0}$) at time $t_{\cal
O}$ is the axial density $\bar s \gamma_4 \gamma_5 d$, then the ratio
of the 3-point correlation function to the two 2-point functions is
$(8/3) B_K$.

$B_K$ is defined to be the value of the matrix element at the physical
kaon and normalized by the Vacuum Saturation Approximation value $ 8/3
M_K^2 F_K^2$
\begin{eqnarray*}
\left\langle K^0 \mathbin\vert Z\ (\bar s d)_{V-A} (\bar s d)_{V-A}(\mu)
                               \mathbin\vert \overline{K^0}   \right\rangle
                 &=&  {(8/3) B_K M_K^2 F_K^2} \,.
\end{eqnarray*}
Earliest calculations of $B_K$ were done using Wilson fermions and
showed significant deviations from this behavior. It was soon
recognized that these lattice artifacts are due to the explicit
breaking of chiral symmetry in the Wilson formulation
\cite{Cabibbo:Bk:PRL1984,Brower:Bk:PRL1984,Bernard:1984bb,Bernard:Bk:1985tm,Gavela:Bk:1988bd}. 
Until 1998, the
only formulation that preserved sufficient chiral symmetry to give the
right chiral behavior was Staggered fermions. First calculations using
this approach in 1989 gave the quenched estimate $B_K (NDR, 2 {\rm
GeV}) = 0.70 \pm 0.01 \pm 0.03$. In hindsight, the error estimates
were highly optimistic, however, the central value was only $10\%$ off
the current best estimate, and most of this difference was due to the
unresolved $O(a^2)$ discretization errors. 

\begin{figure}[tbp]  % 2
\begin{center}
\epsfxsize=0.7\hsize 
\epsfbox{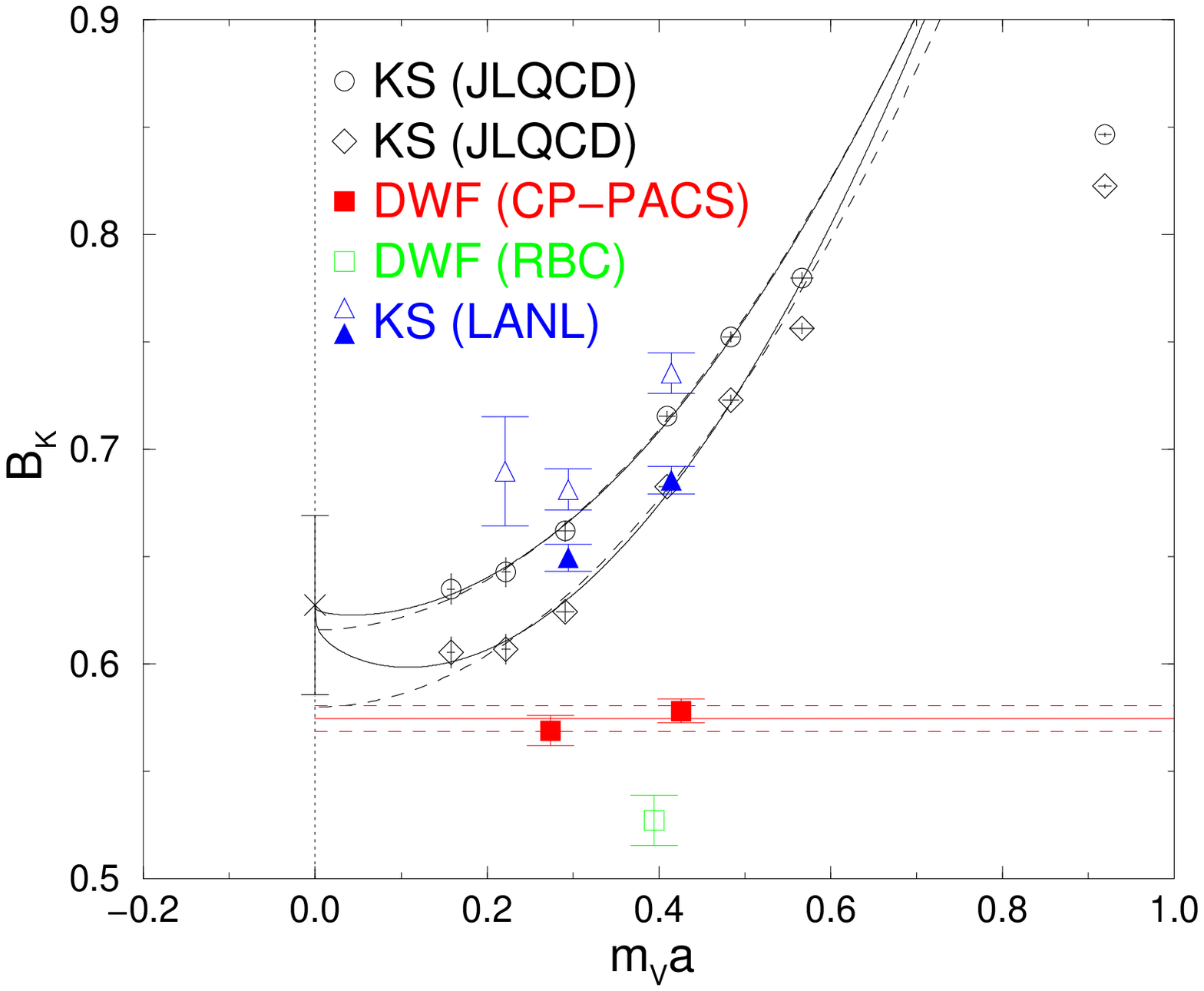}
\end{center}
\caption{Published estimates of $B_K$ with fermion formulations that
respect chiral symmetry. All results are in the quenched
approximation.}
\label{fig:bkestimates}
\end{figure}

In 1997, the staggered collaboration refined its calculation and
obtained $0.62(2)(2)$~\cite{Kilcup:Bk:PRD1998}, again the error
estimate was optimistic as a number of systematic effects were not
fully included. The state-of-the-art quenched calculation using
Staggered fermions was done by the JLQCD collaboration in 1997 and
gave $B_K (2{\rm GeV}) = 0.63 \pm 0.04$~\cite{Aoki:Bk:1998nr}.  This
estimate was obtained using six values of the lattice spacing between
$0.15 - 0.04$ fermi, thus allowing much better control over the
continuum extrapolation as shown in Fig.~\ref{fig:bkestimates} along with other 
published results. This is still the benchmark against which all
results are evaluated and is the value exported to
phenomenologists. This result has three limitations: (i) It is in the
quenched approximation. (ii) All quenched calculations use kaons
composed of two quarks of roughly half the ``strange'' quark mass and
the final value is obtained by interpolation to a kaon made up of
$(m_s/2, m_s/2)$ instead of the physical point $(m_s, m_d)$. Thus,
SU(3) breaking effects ($m_s \neq m_d$) have not been
incorporated. (iii) There are large $O(a^2)$ discretization artifacts,
both for a given transcription of the $\Delta S=2$ operator on the lattice
and for different transcriptions at a given value of the lattice
spacing, so extrapolation to the continuum limit is not as robust as
one would like. These limitations are discussed after a brief summary of 
the recent work. 

%% master copy of fig in ~rajan/papers/epsilon/SLIDES

%
\begin{table}
\begin{center}
\begin{tabular}{|l|c|c|c|c|c|c|c|}
\hline
\multicolumn{1}{|c|}{Collaboration}&
\multicolumn{1} {c|}{year}&
\multicolumn{1} {c|}{$B_K(2 {\rm GeV})$}&
\multicolumn{1} {c|}{Formulation}&
\multicolumn{1} {c|}{Renormalization}&
\multicolumn{1} {c|}{$a^{-1}$ (GeV)} \\
\hline       	     
Staggered~\cite{Kilcup:Bk:PRD1998}  & 1997 & 0.62(2)(2)   & staggered       & 1-loop       &  $\infty$  \\
JLQCD~\cite{Aoki:Bk:1998nr}         & 1997 & 0.63(4)      & staggered       & 1-loop       &  $\infty$  \\
\hline       	     					                                      
Rome~\cite{Becirevic:Bk:2002mm}     & 2002 & 0.63(10)     & Improved Wilson & NP           & $\infty$    \\
Rome~\cite{Becirevic:Bk:2002mm}     & 2002 & 0.70(12)     & Improved Wilson & NP           & $\infty$    \\
\hline       	     					                                      
CP-PACS~\cite{AliKhan:Bk:2001wr}    & 2001 & 0.58(1)      & Domain Wall     & 1-loop       & $1.8$ GeV   \\
CP-PACS~\cite{AliKhan:Bk:2001wr}    & 2001 & 0.57(1)      & Domain Wall     & 1-loop       & $2.8$ GeV   \\
RBC~\cite{Blum:Bk:2001xb}           & 2002 & 0.53(1)      & Domain Wall     & NP           & $1.9$ GeV   \\
\hline    						                                      
DeGrand~\cite{DeGrand:Bk:2002xe}    & 2002 & 0.66(3)      & Overlap         & 1-loop       & $1.6$ GeV   \\
DeGrand~\cite{DeGrand:Bk:2002xe}    & 2002 & 0.66(4)      & Overlap         & 1-loop       & $2.2$ GeV   \\
GGHLR~\cite{GGHLR:Bk:2002}          & 2002 & 0.61(7)      & Overlap         & NP           & $2.1$ GeV  \\
\hline    
\end{tabular}
\end{center}
\caption{Quenched estimates for $B_K$ evaluated in the NDR scheme at
$2 \GeV$. The fermion formulation used in the calculation, the method
used for renormalizing the operators, and the lattice scale at which
the calculation was done are also given. NP indicates non-perturbative
renormalization using the RI/MOM scheme and $a^{-1}=\infty$ implies
that the quoted result is after a continuum extrapolation.}
\label{tab:lattices}
\end{table}

%% master copy of fig in ~rajan/papers/epsilon/SLIDES

In the last four years a number of new methods have been
developed and the corresponding results are summarized in Table 1. 

\begin{itemize}
\item
The Rome collaboration has shown that the correct chiral behavior can
be obtained using $O(a)$ improved Wilson fermions provided
non-perturbative renormalization constants are used. Their latest
results, with two different ``operators'', are $B_K(2{\rm GeV}) =
0.63(10)$ and $0.70(12)$~\cite{Becirevic:Bk:2002mm}. These, while
demonstrating the efficacy of this method, do not supplant the
staggered result, as the continuum extrapolation is based on only
three points and the data have larger errors.  The discretization
errors can be characterized as $B_K(a) = B_K(1 + a \Lambda$) with
$\Lambda \approx 400$ MeV and are similar in magnitude to those with
staggered fermions at $1/a=2$ GeV, as are the differences in estimates
with using different operators.  In the staggered formulation, the
artifacts are, however, $O(a^2 \Lambda^2)$ and $O(\alpha_s^2)$ and the
data suggest an unexpectedly large $\Lambda \sim 900$ Mev.

\item
Four collaborations have new results using domain wall and overlap
fermions as shown in Table~\ref{tab:lattices}
\cite{Blum:Bk:1997mz,Blum:Bk:2001xb,AliKhan:Bk:2001wr,DeGrand:Bk:2002xe,GGHLR:Bk:2002}.
Both formulations have built in chiral symmetry at finite $a$ and
$O(a)$ improvement. Each of these collaborations have used slightly
different methodology, so they cannot be compared head on, or combined
to do a continuum extrapolation.  Thus, the results are quoted with
reference to the lattice spacing at which the calculation was done.
The differences reflect $O(a^2)$ (and $O(\alpha_s^2)$ in cases where
perturbative renormalization constants have been used) artifacts.

\item
Calculations using another method with good chiral behavior, twisted
mass QCD, are in progress~\cite{twistedmass-Bk-lat2001}.

\end{itemize}

Starting with the current best quenched lattice estimate, the JLQCD
staggered result $B_K(2{\rm GeV})= 0.63(4)$, deriving an estimate for
the physical $\hat B_K$ requires consideration of the following
issues.
\begin{itemize}
\item
The $O(a^2)$ errors in the staggered formulation are
large. Nevertheless, the error $0.04$ obtained by the JLQCD
collaboration on including both $O(a^2)$ and $O(\alpha_s^2)$ terms in
the extrapolation is a reasonable $1\sigma$ estimate of both the
statistical and the extrapolation to continuum limit errors.
\item
A choice for $\alpha_s $ and the number of flavors in the perturbative
expression has to be made to convert $B_K \to \hat B_K$. It turns out
that the result is insensitive to whether one uses quenched or full
QCD values. Using the 2-loop expression, the result for the central
value is $\hat B_K = 0.86(6)$.
\item
An estimate of the systematic uncertainty associated with the quenched
approximation and SU(3) breaking.  Preliminary numerical estimates
suggest that dynamical quarks would increase the value by about
$5\%$~\cite{Ishizuka:Bk:1993ya,Kilcup:Bk:1993pa}. Sharpe estimates,
using ChPT, that unquenching would increase $B_K$ by $1.05 \pm 0.15$,
and SU(3) breaking effects would also increase it by $1.05 \pm
0.05$~\cite{sharpe-review-lat1996}.  This analysis of systematic
errors is not robust and, furthermore, the two uncertainties are not
totally independent. So one can take an aggressive and a conservative
approach when quoting the final result for $\hat B_K$.  In the
aggressive approach, the error estimate is given by combining the
central values in quadratures. This gives a $7\%$ uncertainty and
\begin{equation}
\hat B_K = 0.86 \pm 0.06 \pm 0.06 \,.
\end{equation}
In the conservative approach, advocated by
Sharpe~\cite{sharpe-review-lat1996}, one combines the uncertainty in
quadratures to get a $16\%$ uncertainty. The final result in this case is 
\begin{equation}
\hat B_K = 0.86 \pm 0.06 \pm 0.14 \,.
\end{equation}
\end{itemize}

Given the lack of a robust determination of the systematic error, it
is important to decide how to fold these errors in a phenomenological
analysis. The recommendation is to assume a flat distribution for the
systematic error and add to it a gaussian distribution with $\sigma =
0.06$ on either end, and do a separate analysis for the aggressive and
conservative estimates.  In other words, a flat distribution between
$0.72-1.0$ for a conservative estimate of $\hat B_k$ (or $0.80-0.92$
for the aggressive estimate) to account for systematic errors due to
quenching and SU(3) breaking. Since this is the largest uncertainty,
current calculations are focused on reducing it.

Finally, the reasons why the quenched lattice estimate of $B_K$ has
been stable over time and considered reliable within the error
estimates quoted above are worth reemphasizing:
\begin{itemize}
\item
The numerical signal is clean and accurate results are obtained with a 
statistical sample of even 50 decorrelated lattices. 
\item
Finite size effects for quark masses $\ge m_s/2$ are insignificant
compared to statistical errors once the quenched lattices are larger
the $2$ fermi.
\item
In lattice formulations with chiral symmetry, the renormalization constant
connecting the lattice and continuum schemes is small $(< 15\%)$, and
reasonably well estimated by one-loop perturbation theory.
\item
For degenerate quarks, the chiral expansion for the matrix element has
no singular quenched logarithms (they cancel between the $AA$ and $VV$
terms) that produce large artifacts at small quark masses in
observables like $M_\pi^2$, $f_\pi$, etc.  Also, the chiral expansion
between the quenched and full theories have the same form
\cite{Bijnens:CPT:1984ec,Sharpe:QCL:1992ft,Sharpe:TASI:1994dc,Golterman:QCL:1998st}.
\item
ChPT estimates of quenching and SU(3) breaking systematic errors are 
at the $7-16\%$ level~\cite{Sharpe:TASI:1994dc,Ishizuka:Bk:1993ya,Kilcup:Bk:1993pa}.
\end{itemize}

\bibliography{bk}

\printtables
\printfigures
%\printindex

\end{document}